\documentclass[12pt]{article}
\usepackage{graphicx}

\newcommand\pubnumber{SNSN-323-63}
\newcommand\pubdate{\today}

\def\winnipeg{University of Winnipeg \\ Department of Physics\\
515 Portage Avenue, Winnipeg, MB, R3B 2E9, CANADA}
\def\support{\footnote{Work presented for the T2K collaboration.}}

\def\Title#1{\begin{center} {\Large #1 } \end{center}}
\def\Author#1{\begin{center}{ \sc #1} \end{center}}
\def\Address#1{\begin{center}{ \it #1} \end{center}}

\newcommand\pubblock{\rightline{\begin{tabular}{l} \pubnumber\\
         \pubdate  \end{tabular}}}
\newenvironment{Abstract}{\begin{quotation}  }{\end{quotation}}
\newenvironment{Presented}{\begin{quotation} \begin{center} 
             PRESENTED AT\end{center}\bigskip 
      \begin{center}\begin{large}}{\end{large}\end{center} \end{quotation}}
\def\Acknowledgements{\bigskip  \bigskip \begin{center} \begin{large}
             \bf ACKNOWLEDGEMENTS \end{large}\end{center}}

\def\beq{\begin{equation}}
\def\eeq#1{\label{#1}\end{equation}}
\def\eeqn{\end{equation}}

\def\beqa{\begin{eqnarray}}
\def\eeqa#1{\label{#1}\end{eqnarray}}
\def\eeqan{\end{eqnarray}}

\let\bar=\overbar

\def\Dslash{\not{\hbox{\kern-4pt $D$}}}
\def\dslash{\not{\hbox{\kern-2pt $\del$}}}

\def\msb{{\bar{\ssstyle M \kern -1pt S}}}

\begin{document}
\begin{titlepage}
\pubblock

\vfill
\Title{First muon anti-neutrino disappearance oscillation results from T2K }
\vfill
\Author{ Blair Jamieson\support \\
for the T2K Collaboration}
\Address{\winnipeg}
\vfill
\begin{Abstract}

This talk presented the first muon anti-neutrino disappearance analysis
using data from the T2K anti-neutrino data taken in 2014 and up to
March 12, 2015.  The preliminary measured oscillation parameters,
using $2.3\times 10^{20}$ protons on target, are $\Delta
\bar{\rm{m}}_{32}^{2} = 2.33_{-0.23}^{+0.27} \times 10^{-3}$~eV$^{2}$,
and sin$^{2} \bar{\theta}_{23} = 0.515_{-0.095}^{+0.085}$.  These
oscillation parameters are consistent with the neutrino mode, and with
the measurements of the MINOS experiment.

\end{Abstract}
\vfill
\begin{Presented}
Conference on the Intersection of Particle And Nuclear Physics (CIPANP2015)\\
Vail, Colorado,  May 19--24, 2015
\end{Presented}
\vfill
\end{titlepage}
\def\thefootnote{\fnsymbol{footnote}}
\setcounter{footnote}{0}

\section{Introduction}

The Tokai to Kamioka (T2K) is a long baseline neutrino oscillation
experiment situated in Japan\cite{t2k-expt}.  The T2K experiment was
built to measure the accelerator neutrino mixing angle $\theta_{13}$,
to do precision measurements of the atmospheric neutrino mixing
parameters $\theta_{23}$ and $\Delta \rm{m}_{32}^2$, and to search for
a CP-violating phase, $\delta_{CP}$ in the neutrino sector.  Results
of these measurements using a muon neutrino beam have been reported
previously \cite{t2k-joint,t2k-numu, t2k-numu2013, t2k-numu2012, t2k-nue,
  t2k-nue2013, t2k-nue2011}.

The T2K beam is a predominantly muon neutrino beam which is produced
with a high intensity ($\sim$330 kW) 30~GeV proton beam striking a
carbon target at the Japan Proton Accelerator Research Complex
(J-PARC).  The proton beam is measured with a series of beam monitors,
including an Optical Transition Rate (OTR) monitor just upstream of
the target\cite{t2k-otr}.  The positive (negative) charged pions and
kaons produced in this target are focused using a series of three
magnetic hornss.  The positive (negative) charged mesons decay in a
100 m long decay volume producing muon (anti)neutrinos.  Muons making
it to the end of the 100 m long decay pipe are monitored with a Muon
Monitor (MuMon)\cite{t2k-mumon}.  The electron neutrino contamination
in the beam is only $\sim$1\%\cite{t2k-intrinsicnue}\cite{t2k-podnue}.

The beam is measured 280~m from the target by an on-axis detector,
INGRID\cite{ingrid-paper}, which monitors the beam direction, and by
an off-axis detector, ND280.  The tracker section of the magnetized
off-axis detector is comprised of two active scintillator Fine Grained
Detector (FGDs) which act as a neutrino targets\cite{t2k-fgd},
interleaved with three Time Projection Chambers to measure the
momentum and type of charged particles produced in the neutrino
interactions\cite{t2k-tpc}.  Upstream of the tracker is a detector
designed to identify $\pi^0$s produced in neutrino
interactions\cite{t2k-pi0det}.  Surrounding the tracker and $\pi^0$
detectors are electromagnetic calorimeters\cite{t2k-ecal}, and side
muon range detectors between the magnet yokes\cite{t2k-smrd}.

\begin{figure}[htb]
\centering
\includegraphics[width=0.9\textwidth]{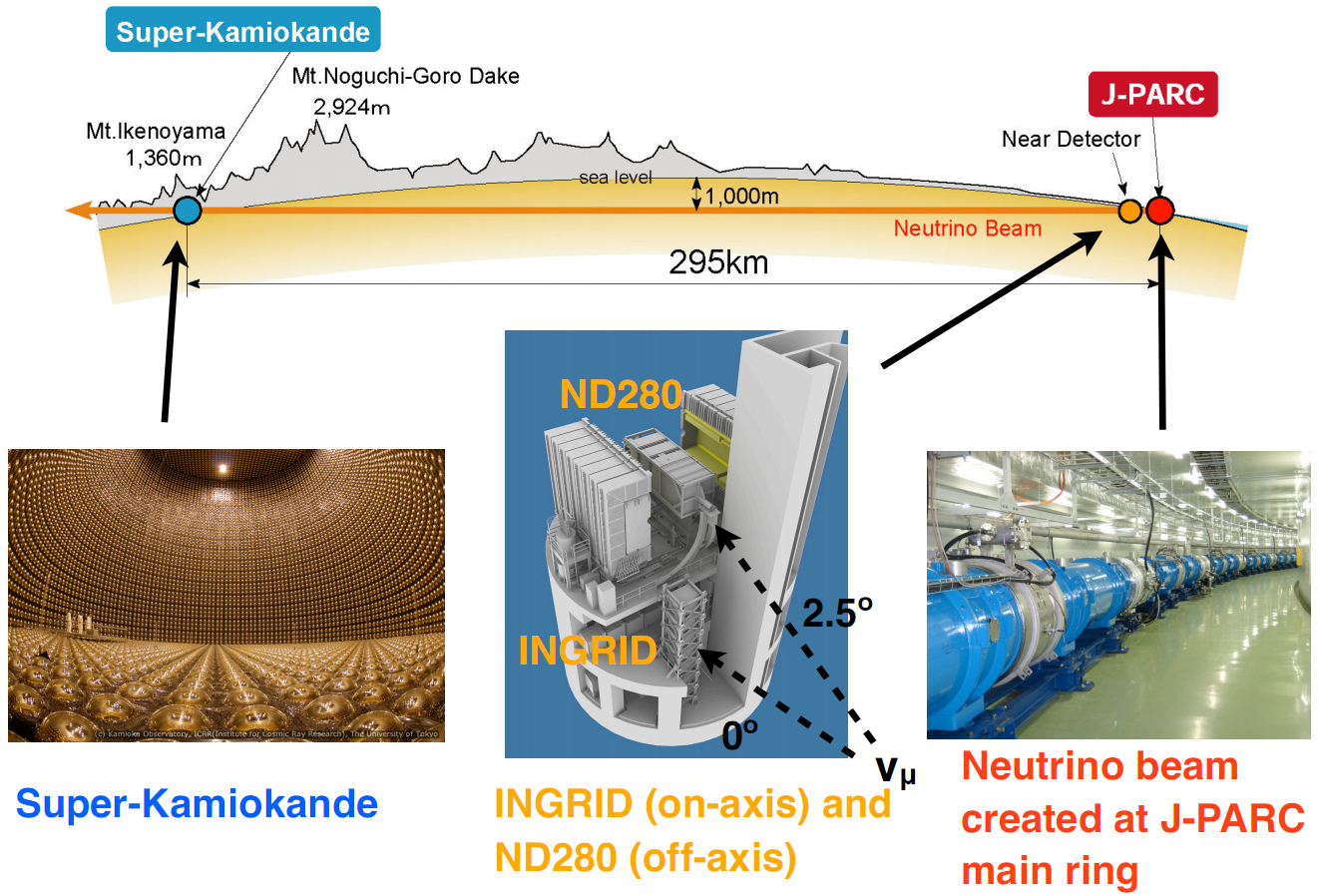}
\caption{Overview of the T2K experiment (top), the T2K far detector
  (bottom-left), the near detectors at 280~m (bottom-middle), and the
  primary proton beamline (bottom-right).}
\label{fig:t2k-overview}
\end{figure}

The T2K far detector, Super-Kamiokande (SK), is 295~km from the
production target, and, similar to ND280, it is off the beam axis by
2.5$^o$.  The off-axis beam is used to tune the energy around
$E\sim0.6$~GeV in order to maximize the neutrino oscillations for the
fixed $L=295$~km to the far detector.  The overview of beam and
detectors of the T2K experiment are shown in
Fig.~\ref{fig:t2k-overview}.

\section{Oscillation analysis flow}

In order to measure the neutrino oscillation parameters $\theta_{23}$
and $\Delta {\rm m}_{32}^2$, the number of $\nu_{\mu}$ events in bins
of reconstructed lepton momentum and angle is predicted using a
detailed simulation.  

\begin{figure}[htb]
\centering
\includegraphics[width=0.48\textwidth]{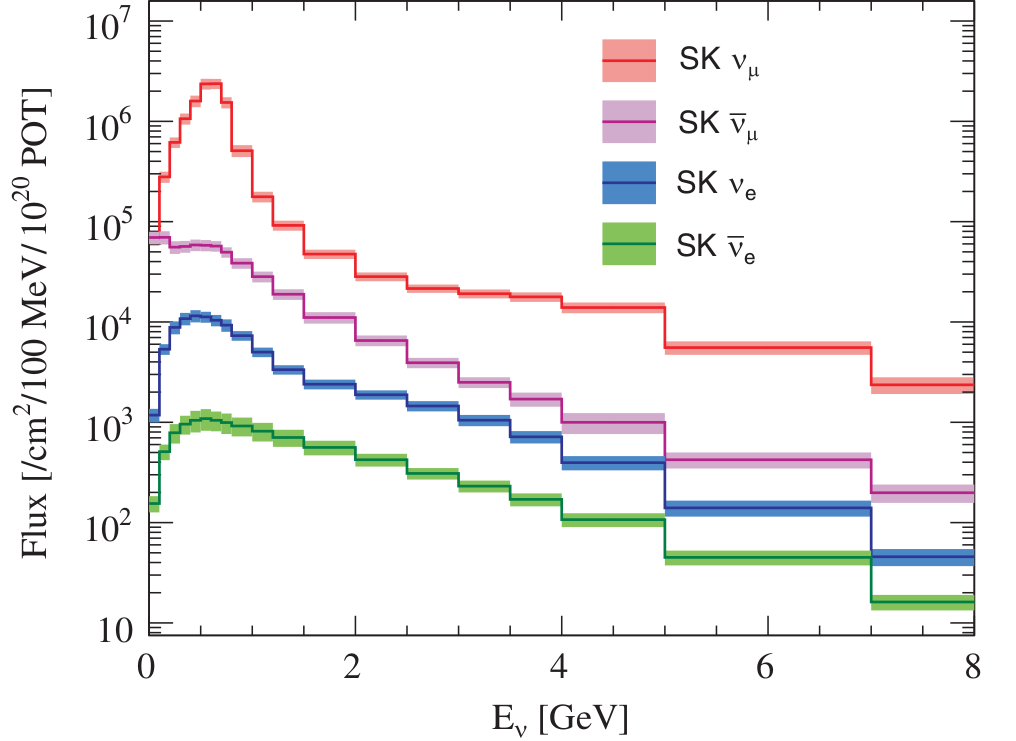}
\includegraphics[width=0.48\textwidth]{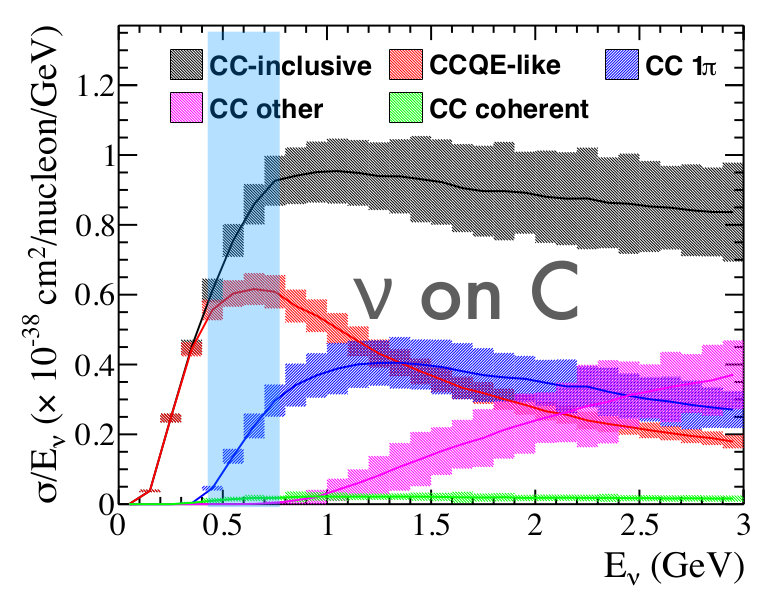}
\caption{The predicted neutrino fluxes at the off-axis near detector,
  ND280, are shown in the left plot, and the neutrino cross sections
  on carbon, before near detector constraints, are shown in the right
  plot.}
\label{fig:fluxxsec}
\end{figure}

Input to the number of events predicted comes from prior global
measurements of the neutrino cross sections, measurement of the pion
and kaon parents to the neutrino production using data from the CERN
NA61 experiment.  This prediction is constrained by a fit to the T2K
near detector data.  For previous analyses of neutrino data, the near
detector $\nu_{\mu}$ interactions were broken into three samples,
based on whether there was no pions, a single pion, or any other
number of particles in the event\cite{t2k-joint}.  For the
anti-neutrino mode, the dataset is smaller, so the near detector data
is only broken into single track and multi-track samples.  The predicted
flux of neutrinos is shown in the left panel of
Fig.~\ref{fig:fluxxsec}, and the predicted cross section is in the
right panel\cite{t2kbeamflux}.  A full covariance matrix of the
anti-correlated uncertainties between the flux and cross section is
used in the event rate prediction at the T2K far detector (Super
Kamiokande).

The oscillation model is described by the
Pontecorvo-Maki-Nakagata-Sakata (PMNS) matrix describing the neutrino
flavour states ($\nu_e$, $\nu_{\mu}$, $\nu_{\tau}$ ) as a quantum
mechanical mixture of the neutrino mass states ($\nu_1$, $\nu_2$,
$\nu_3$).  A full three flavour model, including a CP-violating phase
is used, and includes matter effects.  To first order, ignoring the CP
violation and matter effects, the probability of a $\nu_{\mu}$ of
energy $E$ remaining a $\nu_{\mu}$ after travelling a distance $L$ is:

\begin{equation}
P( \nu_{\mu} \rightarrow \nu_{\mu} ) \approx 1 - {\rm sin}^{2}(
2\theta_{23} ) {\rm sin}^{2}( \Delta m^{2}_{32} \frac{L}{4E} ).
\end{equation}

\section{Neutrino oscillation results with anti-neutrinos}

\begin{figure}[htb]
\centering
\includegraphics[width=0.4\textwidth]{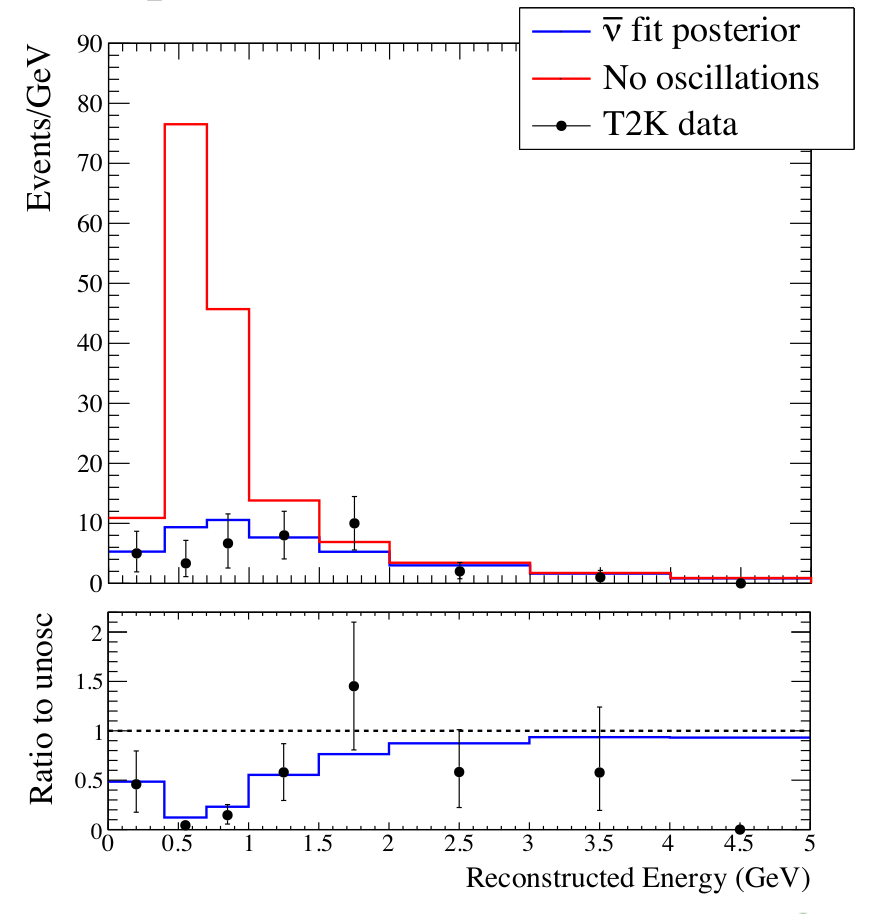}
\includegraphics[width=0.59\textwidth]{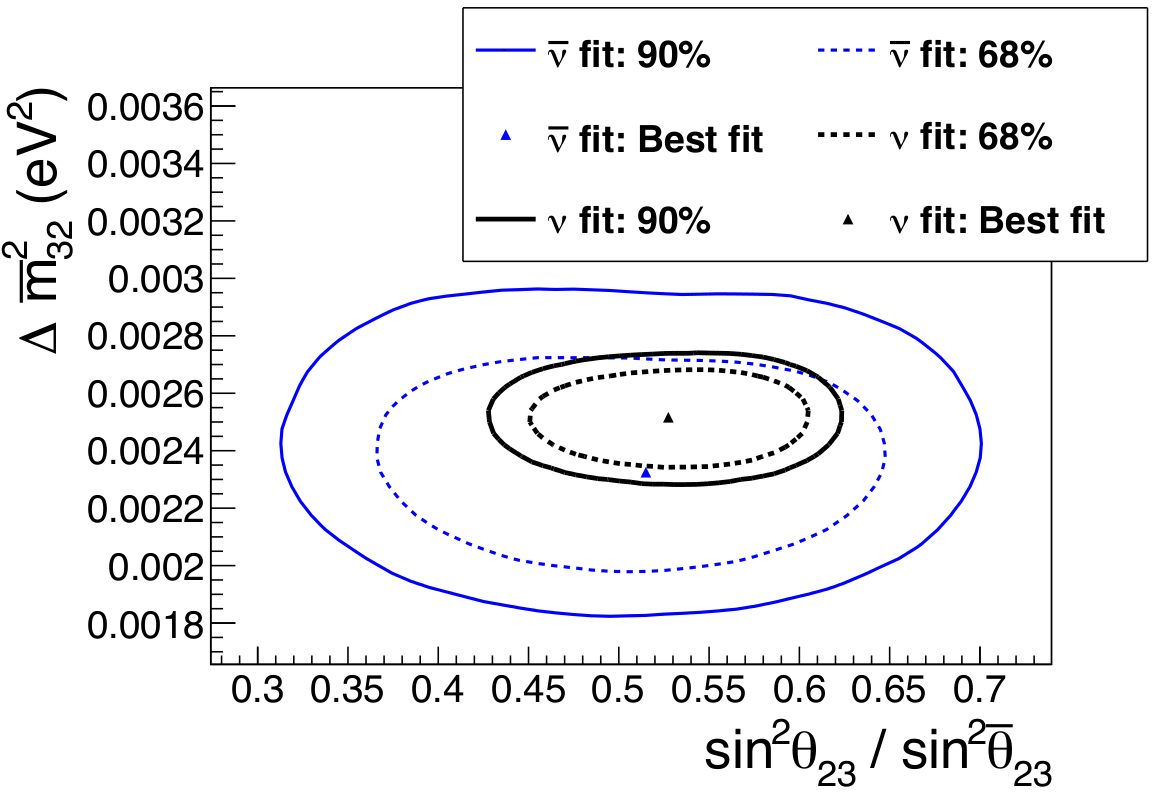}
\caption{The $\bar{\nu}_{\mu}$ data at the T2K far detector are shown
  in the left panels as points, and the predicted number of events is
  shown as the solid lines.  The right panel shows an overlay of the
  best fit $\Delta \rm{m}_{32}^{2}$ and sin$^{2}\theta_{23}$ with 68\%
  and 90\% contours for the anti-neutrino data (blue) and for the
  neutrino data (black).  }
\label{fig:results}
\end{figure}

The muon neutrino event selection in the T2K far detector are events
which are fully contained, have only one reconstructed Cherenkov ring,
have a muon-like particle identification, have momentum greater than
200~MeV/c, and have one or fewer decay electrons.  The predicted
number of events in the far detector for this event selection is 19.9
with oscillation, and 58.9 without oscillation.  The data, shown in
the left panel of Fig.~\ref{fig:results} has 17 candidate
$\bar{\nu}_{\mu}$ events, which clearly favours the oscillation model.

Three separate oscillation analyses were performed, two using
frequentist methods, and the other using Bayesian methods.  In all
cases the fit was evaluated by maximizing a likelihood which is the
product of a Poisson term and systematic uncertainty terms.  In these
fits all of the oscillation parameters were fixed to the PDG2014 and
previous T2K neutrino-mode data values except for the oscillation
parameters $\bar{\theta}_{23}$ and $\Delta \bar{m}_{32}^{2}$.  The
atmospheric neutrino and anti-neutrino oscillation parameters are
treated as independent while the other parameters are the same for
neutrino and anti-neutrino.

The best fit for the oscillation parameters, using $2.3\times 10^{20}$
protons on target, are $\Delta \bar{\rm{m}}_{32}^{2} =
2.33_{-0.23}^{+0.27} \times 10^{-3}$~eV$^{2}$, and sin$^{2}
\bar{\theta}_{23} = 0.515_{-0.095}^{+0.085}$.  These results include
all of the systematic uncertainties, and is still dominated by the
statistical uncertainty.  More data is needed to reach the same level
of precision already achieved with neutrino mode data.  The results
are consistent with the neutrino mode, as shown in the right panel of
Fig.~\ref{fig:results}.  The event prediction with maximal mixing
angle is clearly preferred over the no-oscillation hypothesis.  These
results provide a tighter constraint on the mixing angle than the
previous MINOS results, and are consistent with those
results\cite{MINOS}.

\section{Conclusion}

Using the first anti-neutrino data from the T2K beam, we have shown
that the $\bar{\nu}_{\mu}$ disappearance oscillation is the same as
for $\nu_{\mu}$ within the current statistics.  Results using the
appearance channel are also being analyzed, and will be combined with
the data presented here to put constraints on the CP violating phase
($\delta_{CP}$).  These further results have already been presented at
conferences as these proceedings are being written.

\Acknowledgements

I thank the T2K collaboration for selecting me to present these
results.  We thank the J-PARC staff for superb accelerator performance
and the CERN NA61 collaboration for providing valuable particle
production data. We acknowledge the support of MEXT, Japan; NSERC, NRC
and CFI, Canada; CEA and CNRS/IN2P3, France; DFG, Germany; INFN,
Italy; National Science Centre (NCN), Poland; RSF, RFBR and MES,
Russia; MINECO and ERDF funds, Spain; SNSF and SER, Switzerland; STFC,
UK; and DOE, USA. We also thank CERN for the UA1/NOMAD magnet, DESY
for the HERA-B magnet mover system, NII for SINET4, the WestGrid and
SciNet consortia in Compute Canada, GridPP, UK. In addition
participation of individual researchers and institutions has been
further supported by funds from: ERC (FP7), EU; JSPS, Japan; Royal
Society, UK; DOE Early Career program, USA.

\end{document}